\documentclass[prb,twocolumn,,superscriptaddress]{revtex4-2} 
\usepackage{mathrsfs}
\usepackage{amsfonts}
\usepackage{amsmath}
\usepackage{txfonts}
\usepackage{amssymb}
\usepackage{graphicx,subfigure}
\usepackage{bm}
\usepackage{color}
\usepackage[normalem]{ulem}
\usepackage{dcolumn} 
\usepackage{bbold}
\usepackage[table]{xcolor}

\usepackage[percent]{overpic}

\usepackage{url}
\usepackage[colorlinks=true,
  linkcolor=blue,    
  citecolor=blue,    
  urlcolor=blue      
]{hyperref}

\newcommand{\ket}[1]{|#1\rangle}
\newcommand{\bra}[1]{\langle #1|}

\newcommand{\R}[1]{\textcolor{red}{#1}}

\begin{document}

\title{Inequivalence of Landau-Lifshitz and Landau-Lifshitz-Gilbert dynamics for a 
single quantum spin}
\author{Yuefei Liu}
\affiliation{Department of Physics and Astronomy, Uppsala University, Box 516, 
SE-751 20 Uppsala, Sweden}
\affiliation{Nordita, Stockholm University and KTH Royal Institute of Technology, 
SE-10691 Stockholm, Sweden}
\affiliation{Center for Quantum Spintronics, Department of Physics, Norwegian University 
of Science and Technology NTNU, NO-7491 Trondheim, Norway}

\author{Olle Eriksson}
\affiliation{Department of Physics and Astronomy, Uppsala University, Box 516, 
SE-751 20 Uppsala, Sweden}
\affiliation{WISE-Wallenberg Initiative Materials Science, Uppsala University, Box 516, 
SE-751 20 Uppsala, Sweden}

\author{Erik Sj\"oqvist}
\email{erik.sjoqvist@physics.uu.se}
\affiliation{Department of Physics and Astronomy, Uppsala University, 
Box 516, SE-751 20 Uppsala, Sweden}

\date{\today}

\begin{abstract}
 We examine the relation between the quantum Landau-Lifshitz equation ($q$-LL) 
 [Phys. Rev. Lett. {\bf 110}, 147201 (2013)] and quantum Landau-Lifshitz-Gilbert equation 
 ($q$-LLG) [Phys. Rev. Lett. {\bf 133}, 266704 (2024)]; two non-linear purity preserving 
 master equations that extend classical atomistic spin dynamics into the quantum regime. 
 While the classical LL and LLG counterparts for any number of spins are known to be 
 equivalent, i.e., give identical spin trajectories up to a rescaling of the time parameter, the 
 quantum formulations are equivalent only in certain cases, such as for pure states or for 
 arbitrary single spin-$\frac{1}{2}$ states. Here, we demonstrate that this equivalence 
 breaks down even at the level of a single spin, provided $s \geq 1$. Focusing on a 
 spin-1 particle in an anisotropic crystal field, we show that the $q$-LL and $q$-LLG 
 equations generate inequivalent time evolution. We introduce temporal rescaling 
 misfits that quantify the inequivalence of the two types of dynamics. Although our 
 results highlight fundamental differences in dissipation mechanisms encoded in these 
 equations, the resulting trajectories remain qualitatively similar for this system. 
 \end{abstract}

\maketitle

\section{Introduction}
Atomistic spin dynamics is a computational approach used to simulate and understand 
the magnetic behavior of materials at the atomic scale \cite{antropov96,eriksson17}. In 
atomistic spin dynamics, each atom is assigned a magnetic moment (or site centered 
`atomic spin'), and the time evolution of these spins is calculated using equations 
derived from classical mechanics, such as the Landau-Lifshitz \cite{landau35} or 
Landau-Lifshitz-Gilbert (LLG) \cite{gilbert04} equation. Notably, the LL and LLG 
equations are equivalent, i.e., give identical spin trajectories up to a rescaling of the 
time parameter, for any number of spins \cite{lakshmanan84}. 

More recently, quantum analogs of these equations have been explored. The quantum 
Landau-Lifshitz ($q$-LL) equation, introduced in Ref.~\cite{wieser13}, is formulated as 
a non-linear master equation, with the dissipative `double-bracket' term explicitly involving 
the spin Hamiltonian. In contrast, the dissipative term in the quantum Landau-Lifshitz-Gilbert 
($q$-LLG) equation, proposed in Ref.~\cite{liu24}, takes the form of a `single-bracket' 
expression that involves the density operator and its time derivative.

Although these equations look fundamentally different, as outlined explicitly in the next 
section, they can produce equivalent dynamics in certain special cases. These include 
single spin states that can be expressed in terms of spin operator components, as well 
as arbitrary spin systems in pure state evolution. As discussed in Ref.~\cite{liu24}, the 
equivalence typically breaks down in the case of multi-spin systems in non-pure states. 

In this work, we demonstrate that the dynamics described by the $q$-LL and $q$-LLG 
equations can exhibit fundamental differences, even at the level of a single spin, provided 
$s\geq 1$. To illustrate this inequivalence, we analyze the simplest nontrivial scenario: 
a single spin-1 particle. While the classical LL and LLG equations yield equivalent dynamics, 
the divergent behavior of their quantum counterparts is due to the difference in the underlying 
commutator structure. We identify two primary sources of this inequivalence in the $s=1$ 
system: (i) the presence of an anisotropic crystal field, which breaks the rotational symmetry 
of the spin Hamiltonian, and (ii) the choice of initial quantum state, which can strongly 
influence the spin trajectory under each dynamical framework. These findings highlight 
that the quantum treatment of atomistic spin dynamics introduces qualitatively new behavior, 
even in minimal models, with implications for quantum spin dynamics in magnetic systems. 
Examples of materials relevant for this investigation involve lanthanide impurities (with an 
open 4f shell) in a non-magnetic rare-earth host \cite{pivetta20}, as well as NV centers 
discussed for quantum technologies \cite{jelezko04}. 

\section{Quantum analog spin dynamics}
In the $q$-LL and $q$-LLG formulations, the classical spin vectors are replaced by the 
density operator $\varrho_t$, which encodes the quantum state of the system at time $t$. 
The evolution is described by master equations that combine unitary dynamics (via the 
spin Hamiltonian) with dissipative terms that mimic Gilbert damping. These dissipative 
terms are constructed using non-linear operator bracket structures designed to preserve 
physical properties like positivity and purity conservation.

Explicitly, the $q$-LL and $q$-LLG equations take the form (we put $\hbar = 1$ from 
now on) \cite{wieser13}
\begin{eqnarray}
\dot{\varrho}_t = i[\varrho_t,H] - 
\kappa [\varrho_t,[\varrho_t,H]]  
\label{eq:qll}
\end{eqnarray}
and \cite{liu24}
\begin{eqnarray}
\dot{\varrho}_t = i [\varrho_t,H] + i \kappa [\varrho_t,\dot{\varrho_t}] , 
\label{eq:qllg}
\end{eqnarray}
respectively. Here, $\kappa$ is the Gilbert damping describing the rate of dissipation in the system. 

The key issue we address in this work is the relation between these two equations. Specifically, 
we focus on a single spin in non-pure states. In this case, the two equations are equivalent up 
to a rescaling of time, provided the density operator is of {\it spin-type}, i.e., can be expressed 
in terms of the spin operators ${\bf S} = (S_x,S_y,S_z)$, throughout the time-evolution. In the 
case of $s=1$, such a state takes the form 
\begin{eqnarray}
\varrho_t = \frac{1}{3} \left( \hat{1} + {\bf m}_t \cdot {\bf S} \right) 
\label{eq:spinform}
\end{eqnarray}
with ${\bf m}_t \propto \langle {\bf S} \rangle_t$ corresponding to the magnetization associated 
with the spin. This vector must satisfy $|{\bf m}_t| = |{\bf m}_0| \equiv m_0 \leq 1$ in order for 
the physical requirement $\varrho_t \geq 0$ to hold \cite{remark1}. Note that the purity 
${\rm Tr} \, \varrho_t^2$ increases monotonically with $m_0$, but its maximal value 
${\rm Tr} \, \varrho_t^2 = \frac{5}{9}$ obtained for $m_0 = 1$ corresponds to a non-pure 
state \cite{remark2}. Equation \eqref{eq:qllg} can be turned into the $q$-LL form given 
by Eq.~\eqref{eq:qll} under the rescaling \cite{liu24}
\begin{eqnarray}
t \mapsto \left[1+\frac{1}{9} \kappa^2 m_0^2 \right] t 
\label{eq:scalefactor}
\end{eqnarray}
for dynamics described by Eq.~\eqref{eq:spinform}. 

While a time rescaling is always true for the dynamics of a singe spin-$\frac{1}{2}$, the 
most general $\varrho_t$ for $s\geq 1$ would require the use of a $(2s+1)^2-1$ dimensional 
{\it coherence vector} $\vec{x}_t$ \cite{bertlmann08}. This is vastly different from the restricted 
class of states given by Eq.~\eqref{eq:spinform}, being fully parameterized by the magnetization 
${\bf m}_t \in \mathbb{R}^3$, and this difference would no longer guarantee equivalence 
between the $q$-LL and $q$-LLG dynamics. For a single spin-$1$, the most general state, 
the {\it qutrit-type} state, takes the form  \cite{arvind97} 
\begin{eqnarray}
\varrho_t = \frac{1}{3} ( \hat{1} + \sqrt{3} \vec{x}_t \cdot \vec{\lambda}) 
\label{eq:rho_spin1}
\end{eqnarray}
with $\vec{\lambda} = (\lambda_1,\ldots,\lambda_8)$ a vector of the eight traceless Gell-Mann 
operators (see Appendix \ref{sec:appendixA} for details), and $\vec{x}_t \in \mathbb{R}^8$. 
The coherence vector $\vec{x}_t$ satisfies $0 \leq |\, \vec{x}_t \, |^2 \leq 1$, with 
$|\, \vec{x}_t \, |^2 = 1$ for pure states and $|\, \vec{x}_t \, |^2 = 0$ for the maximally 
mixed state. Note that the purity conservation implies $|\vec{x}_t| = |\vec{x}_0|$. 

\section{Dynamics of a single spin-1}
\subsection{Spin model}
Although the Zeeman term $-\gamma_g {\bf B} \cdot {\bf S}$, $\gamma_g$ the gyromagnetic 
ratio, fully describes the Hamiltonian for isolated spin-$\frac{1}{2}$ particles, material systems 
with $s \geq 1$ can experience nontrivial anisotropy effects that significantly influence the 
dynamics, even for a single spin prepared in a spin-type state (cf. Eq.~\eqref{eq:spinform}). 
Here, we examine $q$-LL and $q$-LLG dynamics of a spin-$1$ system in an anisotropic 
environment that is exposed to a time-independent external magnetic field ${\bf B}$. 

To model this system, we use the Hamiltonian  
\begin{eqnarray}
H = -\gamma_g {\bf B} \cdot {\bf S}  + K_{\perp} S_x^2 + K_{\parallel} S_z^2   
\label{eq:H-spin}
\end{eqnarray}
with ${\bf S} = \left( S_x,S_y,S_z\right)$ spin operators and $K_{\perp},K_{\parallel}$ 
components of the anisotropy tensor along the principal axes. In the following, we take 
$K_{\parallel} < 0 < K_{\perp}$, making the $z$-axis the easy axis along which the 
anisotropy tends to align the spin. Note that the direction of the external magnetic field 
${\bf B}$ does not need to coincide with any of the principal axes.

\subsection{The role of anisotropy}
To elucidate the role of anisotropy for the inequivalence of the $q$-LL and $q$-LLG 
dynamics, we express $H$ in Eq.~\eqref{eq:H-spin} in terms of the eight Gell-Mann 
operators. This is achieved by making use of the relations
\begin{eqnarray}
S_x & = & \frac{1}{\sqrt{2}} 
\left( \lambda_1 + \lambda_6 \right) ,
\nonumber \\ 
S_y & = & \frac{1}{\sqrt{2}} 
\left( \lambda_2 + \lambda_7 \right) , 
\nonumber \\ 
S_z & = & \frac{1}{2} 
\left( \lambda_3 + \sqrt{3} \lambda_8 \right) , 
\label{eq:spinmatrices}
\end{eqnarray}
from which we find   
\begin{eqnarray}
S_x^2 & = & \frac{2}{3} \hat{1} - \frac{1}{4} 
\lambda_3 + \frac{1}{2} \lambda_4 + \frac{1}{4\sqrt{3}} \lambda_8 ,
\nonumber \\ 
S_z^2 & = & \frac{4}{3} \hat{1} + \frac{1}{2} \lambda_3 - \frac{1}{2\sqrt{3}} \lambda_8 .
\label{eq:anisotropymatrices}
\end{eqnarray}
By inserting Eqs.~\eqref{eq:spinmatrices} and \eqref{eq:anisotropymatrices} into 
Eq.~\eqref{eq:H-spin}, we obtain 
\begin{eqnarray}
 & & H = -\frac{\gamma_g B_x}{\sqrt{2}} 
\left( \lambda_1 + \lambda_6 \right) -  \frac{\gamma_g B_y}{\sqrt{2}} \left( \lambda_2 + 
\lambda_7 \right) - \frac{\gamma_g B_z}{2} \left( \lambda_3 + \sqrt{3} \lambda_8 \right) 
\nonumber \\ 
 & & + K_{\perp} \left( - 
\frac{1}{4} \lambda_3 + \frac{1}{2} \lambda_4 + \frac{1}{4\sqrt{3}} \lambda_8 \right) + 
K_{\parallel} \left(  \frac{1}{2} \lambda_3 - \frac{1}{2\sqrt{3}} \lambda_8 \right) , 
\label{eq:H_k}
\end{eqnarray}
where we have omitted unimportant terms that are $\propto \hat{1}$. 

The anisotropy contributions to the Hamiltonian imply that an initial state of the spin-type 
(cf.  Eq.~\eqref{eq:spinform}) would evolve into a qutrit-type state (cf. 
Eq.~\eqref{eq:rho_spin1}), thereby opening up for inequivalent $q$-LL and $q$-LLG 
dynamics. To illustrate this point, we consider the special case where 
$K_{\perp} = 2K_{\parallel}$, for which the Hamiltonian takes the form  
\begin{eqnarray}
H = -\gamma_g {\bf B} \cdot {\bf S} + K_{\parallel} \lambda_4 .
\end{eqnarray}
We calculate the infinitesimal change of the density operator in the $q$-LL case as 
\begin{eqnarray}
\varrho_{\delta t}^{q\text{-LL}} - \varrho_0 & = & 
\Big( -i\gamma_g[\varrho_0,{\bf B} \cdot {\bf S}]
+ iK_{\parallel}[\varrho_0,\lambda_4]
\nonumber\\
 & & + \kappa\gamma_g[\varrho_0,[\varrho_0,{\bf B} \cdot {\bf S}]] - 
 \kappa K_{\parallel}[\varrho_0,[\varrho_0,\lambda_4]]
\Big) \delta t 
\nonumber\\
 & & + \mathcal{O} (\delta t^2),
\label{eq:infiqll} 
\end{eqnarray}
and in the $q$-LLG case as  
\begin{eqnarray}
\varrho_{\delta t}^{q\text{-LLG}} - \varrho_0 & = & \varrho_{\delta t}^{q\text{-LL}} - \varrho_0 
\nonumber \\ 
 & & +\Big( i\kappa^2\gamma_g [\varrho_0,[\varrho_0,[\varrho_0,{\bf B} \cdot {\bf S}]]]
\nonumber \\
 & & -i\kappa^2 K_{\parallel}[\varrho_0,[\varrho_0,[\varrho_0,\lambda_4]]]
\Big)\delta t + \mathcal{O} (\delta t^2) , 
\end{eqnarray}
where we have used Eq.~\eqref{eq:infiqll} to single out the extra contributions in 
the $q$-LLG expression. Suppose now that the system is prepared in the spin-type state 
$\varrho_0 = \frac{1}{3} \left( \hat{1} + m_0 S_z \right)$.  By using the algebra 
of the spin and Gell-Mann operators, one obtains the Zeeman
\begin{eqnarray}
[\varrho_0,[\varrho_0,[\varrho_0,{\bf B}\!\cdot\!{\bf S}]]] 
\propto [\varrho_0,{\bf B}\!\cdot\!{\bf S}].
\end{eqnarray}
and anisotropy  
\begin{eqnarray}
[\varrho_0,[\varrho_0,[\varrho_0,\lambda_4]]] 
\propto  \lambda_5,
\end{eqnarray}
contributions, respectively. Thus, the Zeeman triple-commutator in the $q$-LLG expression  
can be absorbed into the unitary term $i[H,\varrho_0]$ and would therefore just contribute 
to the rescaling of time. In contrast, the anisotropy term generates a new direction 
$\propto \lambda_5$ in SU(3) that makes the change 
$\varrho_0 \mapsto \varrho_{\delta t}^{q\text{-LLG}}$ fundamentally different from that in 
the $q$-LL case. This shows that the two formulations are inequivalent 
as a consequence of the spin anisotropy in this setting.

\begin{figure}[h!]
\centering
\includegraphics[width=\linewidth]{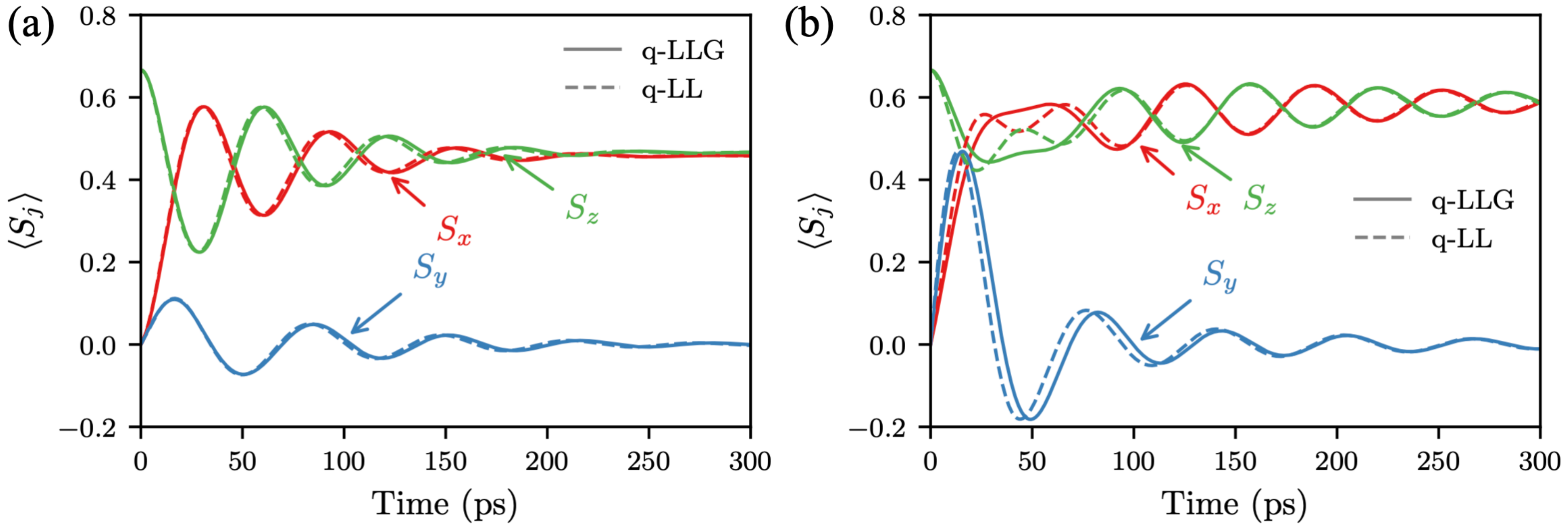}
\caption{(Color online.) The $q$-LLG (solid lines) and $q$-LL (dashed lines) dynamics of 
the spin components. (a) shows the time evolution for the spin-type initial state in 
Eq.~\eqref{eq:spintype_sim} with $m_0=1$ and in the presence of anisotropy [case (i)]. 
(b) shows the time evolution for the qutrit-type initial state in Eq.~\eqref{eq:nonspintype_sim} 
with $p=\frac{5}{6}$ but without anisotropy [case (ii)]. These parameter values are chosen 
so that the two initial states correspond to $\langle {\bf S} \rangle = (0,0,\frac{2}{3})$. An 
oblique magnetic field ${\bf B} = \frac{B_0}{\sqrt{2}}(1,0,1)$ with $|{\bf B}| = B_0$ defining 
the time scale $(\gamma_g B_0)^{-1} \sim 10 {\rm ps}$ and Gilbert damping $\kappa = 0.5$ 
are chosen. The  anisotropy parameters in case (i) are  $K_{\perp}/(\gamma_g B_0) = 0.3$ 
and $K_{\parallel}/(\gamma_g B_0) = -0.1$.}
\label{fig:spin}
\end{figure}

\begin{figure}[h!]
\centering
\includegraphics[width=\linewidth]{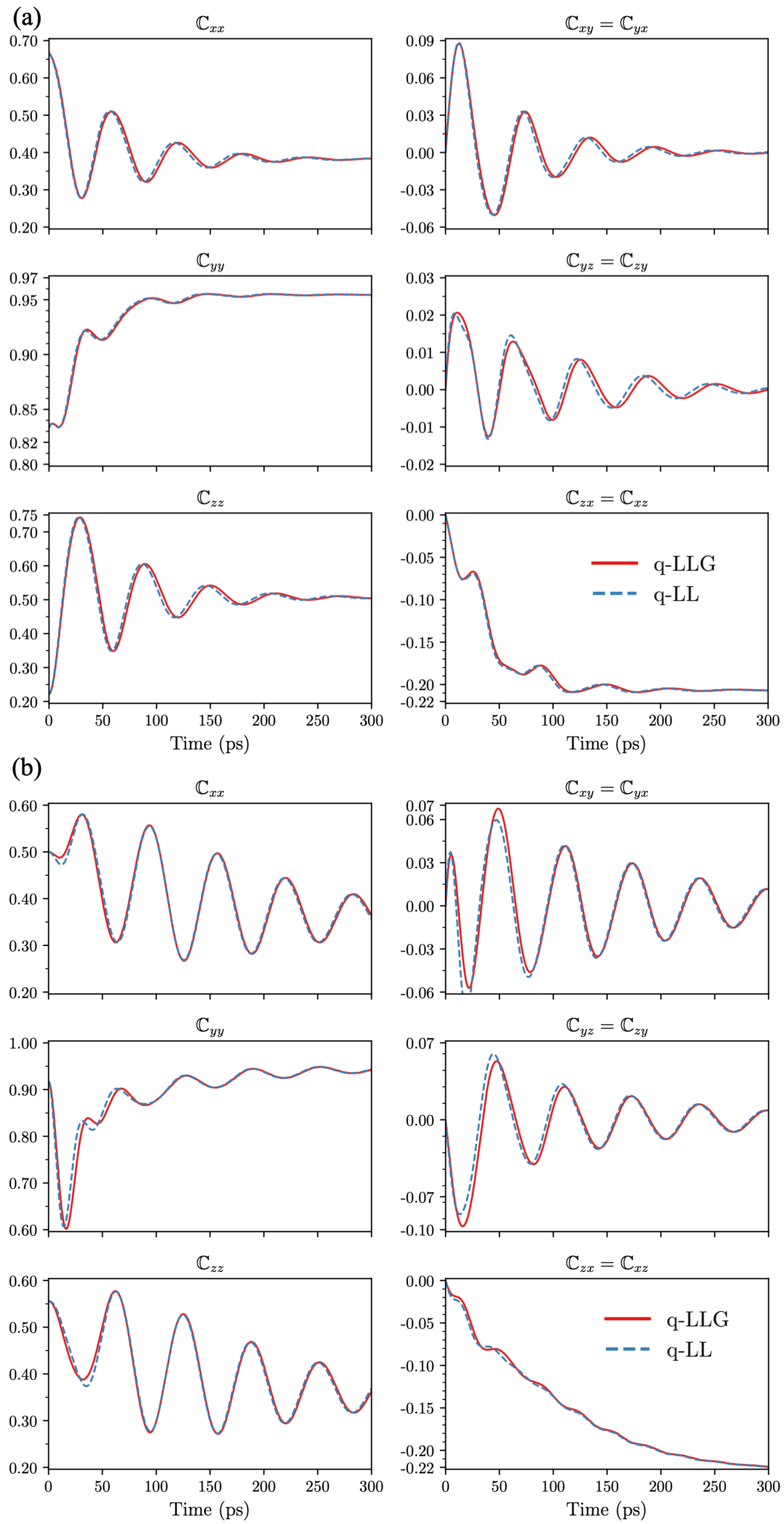}
\caption{(Color online.) The $q$-LLG (solid red lines) and $q$-LL (dashed blue lines) time 
evolution of the covariance matrix elements $\mathbb{C}_{jk} = 
\frac{1}{2} \langle \{ S_{j}, S_{k} \} \rangle - \langle S_{j} \rangle \langle 
S_{k} \rangle$, $j,k=x,y,z$. (a) shows the spin-type initial state in Eq.~\eqref{eq:spintype_sim} 
with $m_0=1$ and in the presence of anisotropy [case (i)]. (b) shows the qutrit-type 
initial state in Eq.~\eqref{eq:nonspintype_sim} with $p=\frac{5}{6}$ but without anisotropy 
[case (ii)]. All numerical values of the dynamical parameters used are identical to those 
in Fig.~\ref{fig:spin}.}
\label{fig:Tjk}
\end{figure}

\begin{figure*}[ht!]
\centering
\includegraphics[width=\linewidth]{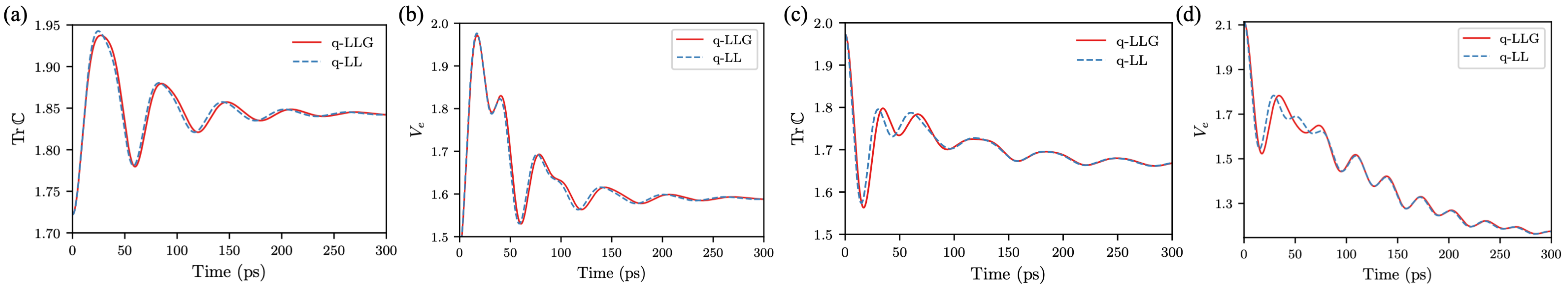}
\caption{(Color online.) The total variance $\big(\Delta S \big)_{\rm tot}^2 (t) = 
\big(\Delta S_x\big)^2 (t) + \big(\Delta S_y\big)^2 (t) + \big(\Delta S_z\big)^2 (t) = 
{\rm Tr} \, \mathbb{C} (t)$ and total quantum fluctuations $V_e (t) = 
\frac{4\pi}{3} \sqrt{\det \, \mathbb{C} (t)}$ under $q$-LLG and $q$-LL dynamics 
of two spin-1 system. (a) and (b) show these quantities for the spin-type initial state 
with anisotropy [case (i)]. (c) and (d) show these quantities for the qutrit-type 
without anisotropy [case (ii)]. All numerical values of the dynamical parameters 
used are identical to those in Fig.~\ref{fig:spin}.}
\label{fig:Ve-TrT}
\end{figure*}

\begin{figure}[ht!]
\centering
\includegraphics[width=0.98\linewidth]{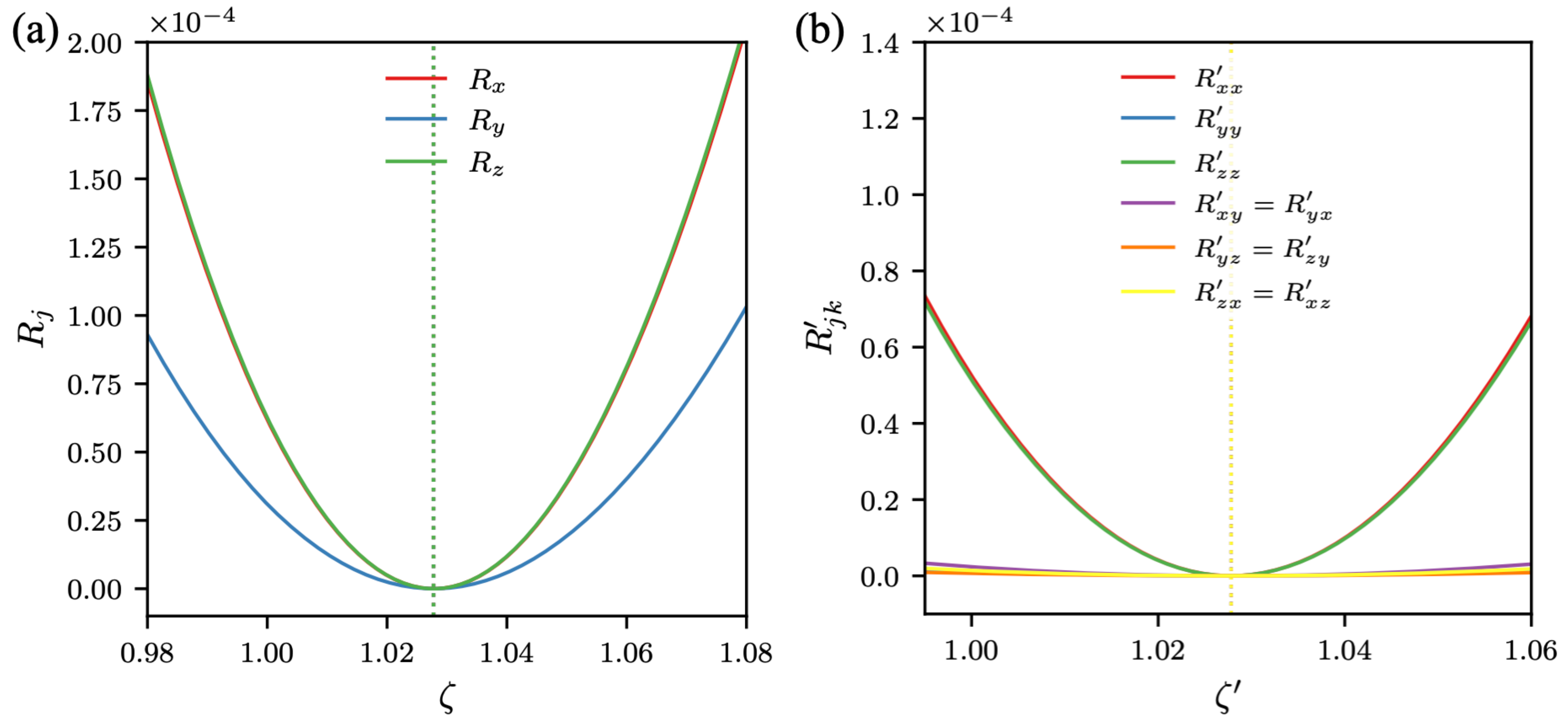}
\caption{(Color online.) The temporal rescaling misfits of (a) spin components and 
(b) covariance matrix elements. Here, we consider a spin-type initial state $m_0 = 1$ 
in Eq.~\eqref{eq:spintype_sim}, vanishing anisotropy, and an oblique magnetic field, 
${\bf B} = \frac{B_0}{\sqrt{2}} (1,0,1)$. The plots confirm the known equivalence 
\cite{liu24} between the $q$-LL and $q$-LLG dynamics with rescaling factor 
$1+\frac{1}{9}0.5^2 \approx 1.028$ in this system.}
\label{fig:rescalable_case}
\end{figure}

\begin{figure}[h!]
\centering
\includegraphics[width=\linewidth]{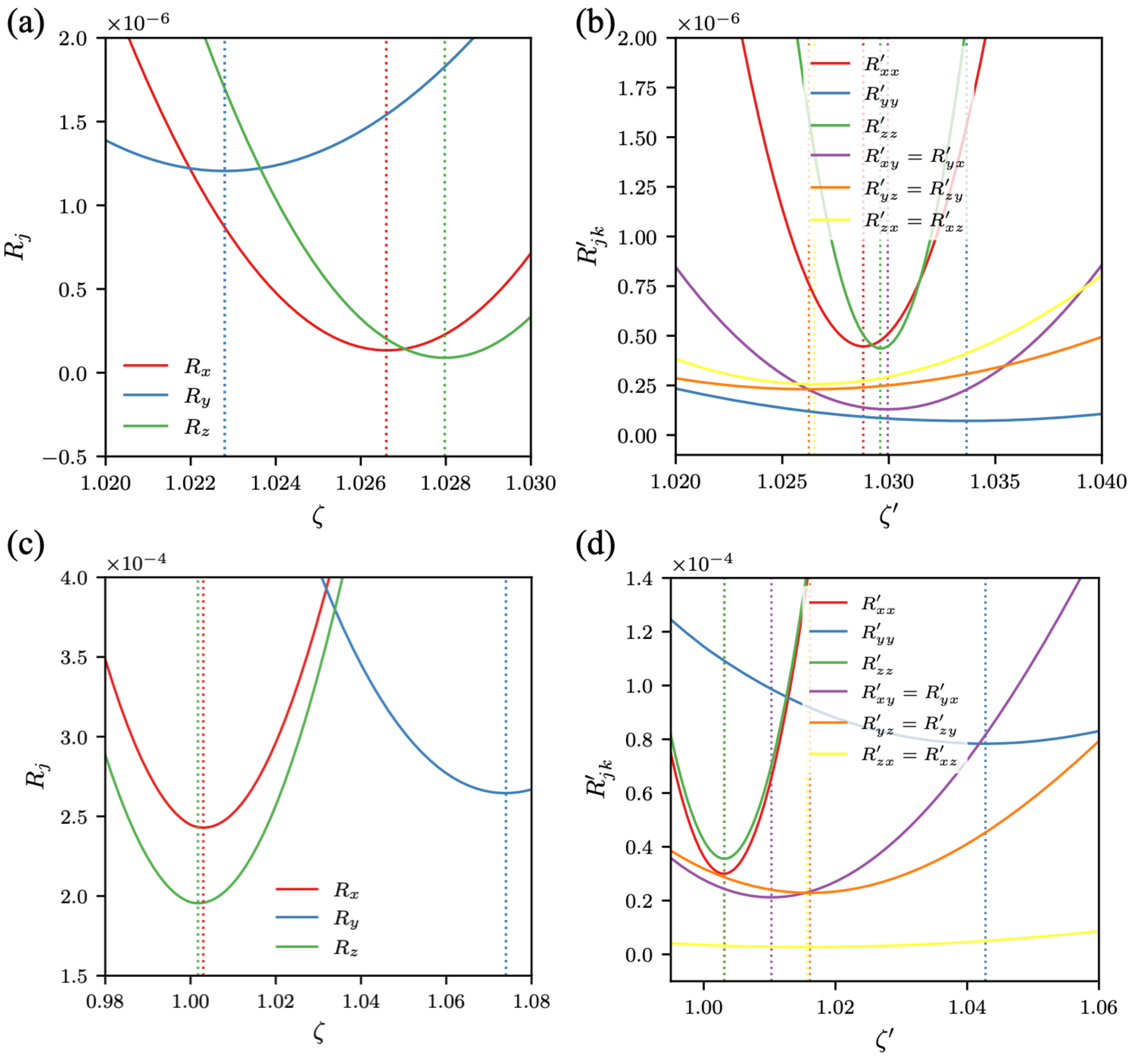}
\caption{(Color online.) The temporal rescaling misfits of spin components and covariance 
matrix elements. The dotted lines mark the mininal values of each time rescaling factors. 
(a) and (b) are temporal rescaling misfits of spin components and covariance matrix 
elements of the spin-type initial state with anisotropy [case (i)]. (c) and (d) are the 
temporal rescaling misfits of spin components and covariance matrix elements of the 
qutrit-type without anisotropy [case (ii)]. The lack of common minima for the various 
vector components and matrix elements demonstrates the inequivalence between 
the $q$-LL and $q$-LLG dynamics. All numerical values of the dynamical parameters 
used are identical to those in Fig.~\ref{fig:spin}.} 
\label{fig:t_rescaling_misfits}
\end{figure}

\begin{figure}
\centering
\includegraphics[width=0.75\linewidth]{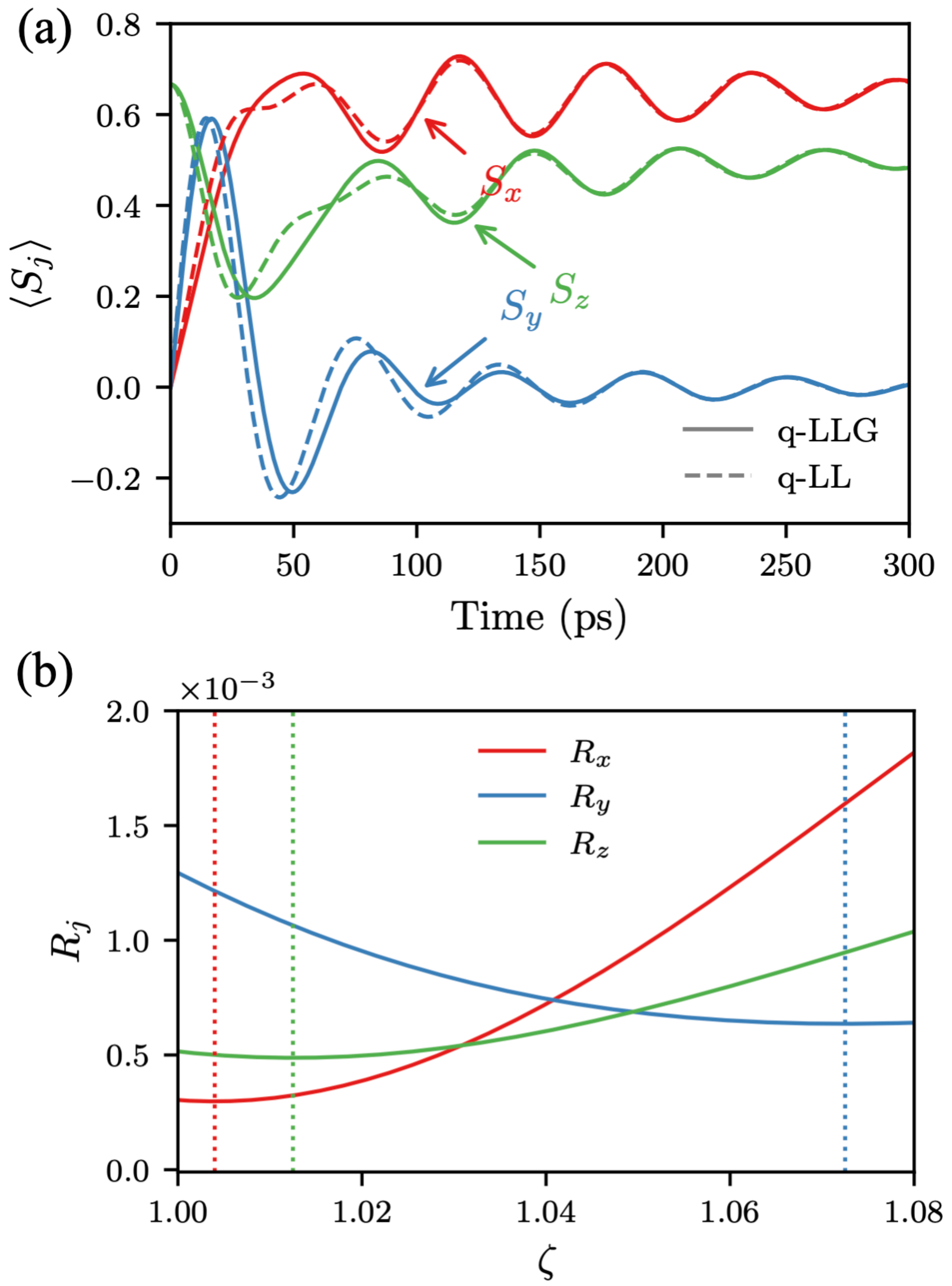}
\caption{(a) The $q$-LLG (solid lines) and $q$-LL (dashed lines) dynamics of the spin components 
for the qutrit-type initial state in Eq.~\eqref{eq:nonspintype_sim} with $p=\frac{5}{6}$ and 
non-vanishing anisotropy. (b) The  combined qutrit-type initial state and non-vanishing 
anisotropy, results in temporal rescaling misfits $R_j$ which are larger than those shown in 
Fig.~\ref{fig:t_rescaling_misfits} (c) by almost one order of magnitude.  All numerical values 
of the dynamical parameters used are identical to those in Fig.~\ref{fig:spin}.}
\label{fig:nonSpin_withKs}
\end{figure}

\subsection{Numerical analysis}
To demonstrate the inequivalence of the $q$-LL and $q$-LLG dynamics for the spin-$1$ 
system, we numerically compare them for specific choices of parameters and initial states. 
To fully characterize the spin state in the simulations, we use the spin components 
$\langle S_x \rangle,\langle S_y \rangle,\langle S_z \rangle$, as well as the covariance 
matrix $\mathbb{C}$ defined as \cite{bharath19} 
\begin{eqnarray}
\mathbb{C}_{jk} & = & \frac{1}{2} \left\langle 
\left\{ S_{j} - \langle S_{j} \rangle ,S_{k} - \langle S_{k} \rangle \right\} \right\rangle 
\nonumber \\ 
 & = & \frac{1}{2} \left\langle \left\{ S_{j}, S_{k} 
\right\} \right\rangle - \langle S_{j} \rangle  \langle S_{k} \rangle , \quad j,k=x,y,z .
\label{eq:covariance_matrix}
\end{eqnarray} 
While the former defines the magnetization associated with the spin, the latter tracks 
the quantum fluctuations along the evolution path. 

The covariance matrix can be pictured as an ellipsoid surrounding the tip of the evolving 
spin vector $\langle {\bf S} \rangle (t) = {\rm Tr} ({\bf S} \varrho_t )$. The direction and 
length of the ellipsoid's half axes are the normalized eigenvectors and squared eigenvalues, 
respectively, of the covariance matrix. Thus, the  volume of the ellipsoid is given by 
$V_e (t) = \frac{4 \pi}{3} \sqrt{{\rm det}\, \mathbb{C} (t)}$. An important feature of 
$V_e (t)$ is that it vanishes when the spin state has no uncertainty (zero variance) in 
one direction. This makes $V_e (t)$ a useful scalar quantifier of total uncertainty and 
correlations among the spin components: larger determinant means greater total 
quantum fluctuations in all directions. A second useful quantity is ${\rm Tr} \, \mathbb{C} (t)$, 
which is the total variance $(\Delta S)^2_{\text{tot}} (t) = \left( \Delta S_x \right)^2 (t) + 
\left( \Delta S_y \right)^2 (t) + \left( \Delta S_z \right)^2 (t)$ with $\Delta S_j^2 (t) = 
{\rm Tr} \left( S_j^2 \varrho_t \right) - {\rm Tr} \left[ \left( S_j \varrho_t \right) \right]^2$, $j=x,y,z$. 
This quantity satisfies $1 \leq (\Delta S)^2_{\text{tot}} \leq 2$, with lower bound 
corresponding to a `quasi-classical' spin coherent state \cite{peres95} and the upper 
bound any state for which $\langle {\bf S} \rangle$ vanishes (for further details, see 
Appendix \ref{sec:appendixB}). 

We will now numerically examine two distinct cases: 
\begin{enumerate}
\item[(i)] The spin is initially prepared in the spin-type state 
\begin{eqnarray}
\varrho_0 = \frac{1}{3} \left(\hat{1} + m_0 S_z\right)
\label{eq:spintype_sim}
\end{eqnarray}
and evolves under the influence of anisotropy and an applied field.
\item[(ii)] The spin is prepared in a qutrit-type state 
\begin{eqnarray}
\varrho_0 & = & p \ket{1;1} \bra{1;1} + (1-p) \ket{1;-1} \bra{1;-1}  
\nonumber \\ 
 & = & \frac{1}{3} \left( \hat{1} + (3p - 2) S_z + \lambda_3 \right) , 
 \label{eq:nonspintype_sim}
\end{eqnarray}
and evolves in the absence of anisotropy but an applied field. 
\end{enumerate}
For case (i), we choose $m_0 = 1$, corresponding to a non-pure spin-type state for which 
the purity takes the maximal value ${\rm Tr} \varrho_0^2 = \frac{5}{9}$ \cite{remark2}. We 
further use the anisotropies $K_{\perp}/(\gamma_g B_0) = 0.3$ and 
$K_{\parallel}/(\gamma_g B_0) = -0.1$. For case (ii), we choose $p=\frac{5}{6}$, and put 
$K_{\perp} = K_{\parallel} =0$. In both simulations, we choose the precession time-scale 
$(\gamma_g B_0)^{-1} \sim 10$ ps, and Gilbert damping $\kappa = 0.5$. The external 
magnetic field is taken to be ${\bf B} = \frac{B_0}{\sqrt{2}} (1,0,1)$, i.e., an oblique field 
at $45^{\circ}$ to $z$-axis. The two cases (i) and (ii) are complementary in the sense that 
the former shows the effect of anisotropy and the latter shows the effect of the initial state 
on the inequivalence between the $q$-LL and $q$-LLG dynamics. While the parameters 
$m_0$ and $p$ are chosen so that the two initial states correspond to 
$\langle {\bf S} \rangle = (0,0,\frac{2}{3})$, the corresponding covariance matrices 
differ at $t=0$. 

Figure~\ref{fig:spin} shows the time evolution of the spin vector components 
$\langle S_j \rangle$, $j=x,y,z$, and Fig.~\ref{fig:Tjk} shows the corresponding covariance 
matrix elements, in the case of $q$-LLG and $q$-LL dynamics. In addition, the total 
uncertainty and total variance of the spin-1 states obtained by solving the two equations 
are shown in Fig.~\ref{fig:Ve-TrT}. Note that two different starting states are considered 
in these figures. The plots in Figs.\ref{fig:Tjk} and \ref{fig:Ve-TrT} show that while these 
two types of dynamics match closely, despite the relatively large Gilbert damping, one 
can discern their inequivalence in the details.  

To quantitatively confirm that the dynamics of $q$-LLG and $q$-LL are indeed inequivalent 
under time rescaling, we introduce and numerically compute the {\it temporal rescaling misfits} 
(from $q$-LLG to $q$-LL) of the spin vector components, defined as       
\begin{eqnarray}
R_j(\zeta) = \frac{1}{N_{\rm grid}} \sum_{t} \Bigl[S_{j}^{(\text{$q$-LLG})}( {\zeta}t)  
\;-\;  \,S_{j}^{(\text{$q$-LL})}( t) \Bigr]^{2},
\label{eq:t-rescaling-misfits-spin}
\end{eqnarray}
as well as of the covariance matrix elements:  
\begin{eqnarray}
R_{ij}'(\zeta') &=&
\frac{1}{N_{\rm grid}} \sum_{t} \Bigl[ \mathbb{C}_{ij}^{(\text{$q$-LLG})}({\zeta'}t)
\;-\; \mathbb{C}_{ij}^{(\text{$q$-LL})}( t)
\Bigr]^{2} ,
\label{eq:t-rescaling-misfits-T-ij}
\end{eqnarray}
both summing over the time grid ($N_{\rm grid} = 5\cdot 10^4$ equidistant grid points). 
Here, $\zeta$ and $\zeta'$ are temporal scale factors. For the two types of evolution to 
be equivalent, it must hold that all minimal temporal rescaling factors $\zeta,\zeta'$ are 
the same for all vector components $\langle S_j \rangle$ and covariance matrix elements 
$\mathbb{C}_{jk}$. Figure \ref{fig:rescalable_case} shows the rescaling misfit plots for 
vanishing anisotropy and spin-type initial state, confirming the known equivalence 
\cite{liu24} between $q$-LL and $q$-LLG dynamics with a rescaling factor 
$1+\frac{1}{9}0.25^2 \approx 1.028$ (cf. Eq.~\eqref{eq:scalefactor} with $m_0=1$). 
On the other hand, Fig.~\ref{fig:t_rescaling_misfits} shows that the minimal temporal 
rescaling factors are different for the various components of $\langle {\bf S} \rangle$ 
and elements of $\mathbb{C}$. In other words, there is no uniform time rescaling, 
neither for the spin vector, nor for the covariance matrix, despite the initial state 
being of spin-type or in the absence of spin anisotropy. These two examples for 
a single spin-$1$, is a direct proof of $q$-LLG and $q$-LL equations are indeed 
two inequivalent dynamical processes, even though the overall behavior of the 
spin is very similar in the two descriptions.

To further test the above noted similarity, we simulate a less symmetric setting, where the 
qutrit-type state in Eq.~\eqref{eq:nonspintype_sim} with $p=\frac{5}{6}$ is combined with 
non-vanishing anisotropy. Figure \ref{fig:nonSpin_withKs} shows the spin components 
and the corresponding temporal rescaling misfits. Although, the difference between the 
two types of dynamics is more clear than in Figs.~\ref{fig:spin} and \ref{fig:t_rescaling_misfits}, 
the overall behavior remains similar. 

\section{Conclusions}
We have examined the $q$-LL and $q$-LLG dynamics of a single spin-1 particle in an 
anisotropic crystal environment and found that the two frameworks are fundamentally 
inequivalent. Specifically, the time evolution generated by $q$-LL and $q$-LLG cannot 
be related by any simple rescaling of time. This is in contrast to the classical atomistic 
spin dynamics framework, where the Landau–Lifshitz and Landau–Lifshitz–Gilbert 
equations are equivalent up to time rescaling for any number of spins. 

The inequivalence highlights a distinct non-classical feature of quantum spin dynamics 
that emerges even in the minimal setting of a single spin with $s=1$. Despite their 
inequivalence, both the $q$-LL and $q$-LLG dynamics appear to produce similar 
trajectories in the system considered here. This similarity has been seen also in the 
multispin-$\frac{1}{2}$ cases reported in Ref.~\cite{mirzaei25}. This suggests that, 
although the underlying commutator structures differ, the physical predictions closely 
match; an observation that warrants further investigation in other spin models.

\section*{Acknowledgments}
Y.L. also acknowledges the financial support from Nordita, Grant No. ERC-SYG 81451.
O.E. acknowledges financial support from the Swedish Research Council (VR) and the 
Knut and Alice Wallenberg foundation (KAW). O.E. also acknowledges support from the 
Wallenberg Initiative Materials Science (WISE), funded by the Knut and Alice Wallenberg 
Foundation, for support, as well as support from STandUPP, the ERC (FASTCORR project) 
and eSSENCE. E.S. acknowledges financial support from the Swedish Research Council 
(VR) through Grant No. 2025-05249.

\appendix 
\section{Spin-$1$ and Gell-Mann operators}
\label{sec:appendixA}
In terms of the standard spin-$1$ basis $\{ \ket{1;1}, \ket{1;0}, \ket{1;-1} \}$, the spin operators read 
\begin{eqnarray}
S_x & = & \frac{1}{\sqrt{2}} \left( \ket{1;1} \bra{1;0} + \ket{1;0} \bra{1;-1} + {\rm H.c.} \right), 
\nonumber \\ 
S_y & = & \frac{1}{\sqrt{2}} \left(  -i \ket{1;1} \bra{1;0} -i \ket{1;0} \bra{1;-1} + {\rm H.c.} \right) , 
\nonumber \\ 
S_z & = & \ket{1;1} \bra{1;1} - \ket{1;-1} \bra{1;-1} . 
\end{eqnarray}
Similarly, the Gell-Mann operators are taken to be   
\begin{eqnarray}
\lambda_1 & = & \ket{1;1} \bra{1;0} + \ket{1;0} \bra{1;1} ,
\nonumber \\ 
\lambda_2 & = & -i\ket{1;1} \bra{1;0} + i\ket{1;0} \bra{1;1} ,
\nonumber \\ 
\lambda_3 & = & \ket{1;1} \bra{1;1} - \ket{1;-1} \bra{1;-1} , 
\nonumber \\
\lambda_4 & = & \ket{1;1} \bra{1;-1} + \ket{1;-1} \bra{1;1} , 
\nonumber \\
\lambda_5 & = & -i \ket{1;1} \bra{1;-1} + i \ket{1;-1} \bra{1;1} , 
\nonumber \\
\lambda_6 & = & \ket{1;0} \bra{1;-1} + \ket{1;-1} \bra{1;0} , 
\nonumber \\
\lambda_7 & = & -i \ket{1;0} \bra{1;-1} + i \ket{1;-1} \bra{1;0} , 
\nonumber \\
\lambda_8 & = & \frac{1}{\sqrt{3}} \big( \ket{1;1} \bra{1;1} + \ket{1;0} \bra{1;0} 
\nonumber \\ 
 & & - 2\ket{1;-1} \bra{1;-1} \big) . 
\end{eqnarray}

\section{Upper and lower bound of total variance}
\label{sec:appendixB}
First note that  
\begin{eqnarray}
(\Delta S)^2_{\text{tot}} = \langle {\bf S}^2 \rangle - \left| \langle {\bf S} \rangle \right|^2.   
\end{eqnarray}
We recall that for a spin-$1$ system,
\begin{eqnarray}
{\bf S}^2 & = & S_x^2 + S_y^2 + S_z^2 = 2\hat{1} 
\nonumber \\ 
 & \Rightarrow & \langle {\bf S}^2 \rangle = 2.   
\end{eqnarray}
In other words, the upper and lower bounds of the total variance is completely determined by 
$\left| \langle {\bf S} \rangle \right|$. This quantity satisfies the inequality
\begin{eqnarray}
0 \leq \left| \langle {\bf S} \rangle \right| \leq 1.
\end{eqnarray}
The lower bound is obtained, e.g., by a random mixture $\frac{1}{3} \hat{1}$ or a pure 
`anti-coherent' state \cite{zimba06} that can take the form 
$a\ket{1;1} + b\ket{1;0} - a\ket{1;-1}$, $a,b$ real numbers such that $2a^2+b^2=1$. 
The upper bound characterizes a spin-coherent state \cite{peres95}, which is a pure 
state with maximal projection along some arbitrary direction. We thus conclude that 
\begin{eqnarray}
1 \leq (\Delta S)^2_{\text{tot}} \leq 2.
\end{eqnarray}

\end{document}